\documentclass[preprint]{sigplanconf}

\usepackage{tikz}
\usetikzlibrary{positioning}

\usepackage{amsmath}
\usepackage{mathrsfs}

\usepackage{listings}
\usepackage{breqn}
\usepackage{graphicx}

\newcommand{\overbar}[1]{\mkern 1.5mu\overline{\mkern-1.5mu#1\mkern-1.5mu}\mkern 1.5mu}

\usepackage{courier}            
\usepackage[scaled]{helvet} 
\usepackage{xspace}
\usepackage{proof}
\usepackage{graphicx}
\usepackage{amsmath}
\usepackage{amsthm}
\usepackage{amscd}
\usepackage{amssymb}
\usepackage{float}
\usepackage{epsfig}
\usepackage{xcolor}
\usepackage{color}

\usepackage[T1]{fontenc}
\usepackage[kerning,spacing]{microtype}
 
\newcommand{\hide}[1]{} 
\newcommand{\C}[1]{\lstinline!#1!}

\newcommand{\figlabel}[1]{\label{fig:#1}}
\newcommand{\longfigref}[1]{Figure~\ref{fig:#1}}

\newcommand{\figref}{\longfigref}

\numberwithin{equation}{section}

\usepackage[T1]{fontenc}

\definecolor{dkgreen}{rgb}{0,0.3,0}
\definecolor{gray}{rgb}{0.5,0.5,0.5}
\definecolor{mauve}{rgb}{0.58,0,0.82}
\definecolor{light-gray}{gray}{0.80}

\usepackage{listings}
\lstdefinelanguage{sketch}{
  morekeywords = {
       bool, harness, data, int, bool, adt, new, return, assert , assume, let, case, switch},
  morecomment=[s]{/*}{*/},
}

\renewcommand{\scriptsize}{\fontsize{8.5}{9}\selectfont}
\makeatletter
\lst@AddToHook{TextStyle}{\let\lst@basicstyle\scriptsize\fontfamily{lmss}\selectfont}
\makeatother

\usepackage{url}

\renewcommand{\vss}{\vspace{10pt}}

\usepackage{rotating}

\usepackage{enumitem}

\newcommand{\conf}[1]{}

\begin{document}

\setlength{\pdfpageheight}{\paperheight}
\setlength{\pdfpagewidth}{\paperwidth}

\conferenceinfo{CONF 'yy}{Month d--d, 20yy, City, ST, Country}
\copyrightyear{20yy}
\copyrightdata{978-1-nnnn-nnnn-n/yy/mm}
\copyrightdoi{nnnnnnn.nnnnnnn}


\titlebanner{}        
\preprintfooter{}   

\title{sk\_p: a neural program corrector for MOOCs}

\authorinfo{Yewen Pu}
           {MIT}
           {yewenpu@mit.edu}
\authorinfo{Karthik Narasimhan}
           {MIT}
           {karthikn@mit.edu}
\authorinfo{Armando Solar-Lezama}
           {MIT}
           {asolar@csail.mit.edu}
\authorinfo{Regina Barzilay}
           {MIT}
           {regina@csail.mit.edu}



\maketitle

\begin{abstract}
We present a novel technique for automatic program correction in MOOCs, capable of fixing both syntactic and semantic errors without manual, problem specific correction strategies. Given an incorrect student program, it \emph{generates} candidate programs from a distribution of likely corrections, and checks each candidate for correctness against a test suite.

The key observation is that in MOOCs many programs share similar code fragments, and the seq2seq neural network model, used in the natural-language processing task of machine translation, can be modified and trained to recover these fragments.

Experiment shows our scheme can correct 29\% of all incorrect submissions and out-performs state of the art approach which requires manual, problem specific correction strategies. 
\end{abstract}




\section{Introduction}

Massive open online courses (MOOCs) have become a popular way of teaching programming. According to one ranking, 10 of the top 20 most popular MOOCs are in computer science, and several of these are introductory programming courses\footnote{http://www.onlinecoursereport.com/the-50-most-popular-moocs-of-all-time/}. An important problem for MOOCS that aim to teach programming is providing accurate feedback to students when they make programming mistakes; this is particularly important for introductory programming courses where students are not yet proficient with debugging techniques.


Prior work based on symbolic program synthesis has demonstrated that with some instructor guidance, it is feasible to provide this feedback for small introductory programming assignments~\cite{rishabh}. This prior approach, however, comes with significant caveats. First, it requires the instructor to provide an \emph{error model} that describes the space of corrections that the system can explore.
Writing an error model that is sufficiently detailed to correct a large fraction of submissions but limited enough to allow for short correction times is not easy to do.
Second, the reliance on symbolic analysis makes the system brittle; assignments need to adhere to the subset of python modeled by the system, and student programs must be syntactically correct for the system to even attempt a correction.

This paper follows a different approach that is based on the idea of \emph{data-driven synthesis} (DDS), which has recently been applied successfully in domains including program repair~\cite{fan}, inferring program properties ~\cite{raychev2},  and program completion~\cite{raychev}. The general framework of DDS is illustrated in \figref{ddsfigure}. In this framework, a learning algorithm is used during training time to produce a model of the problem at hand. Given an incomplete or erroneous program (the seed program), this model can produce a distribution of candidate completions or corrections. This distribution is used by a synthesis algorithm to find candidate solutions that have high probability according to the model and also are correct according to a potentially incomplete specification. DDS is particularly well suited to our problem because (a) given the scale of a MOOC, one can get a large corpus of solutions to \emph{the exact same assignment}, allowing us to train very accurate models. Additionally, (b) in this domain it is already customary to define the correctness of a submission in terms of a rich hand-crafted test suite, which can serve as a very strong specification for the DDS system.
\begin{figure}
\includegraphics[width=\columnwidth]{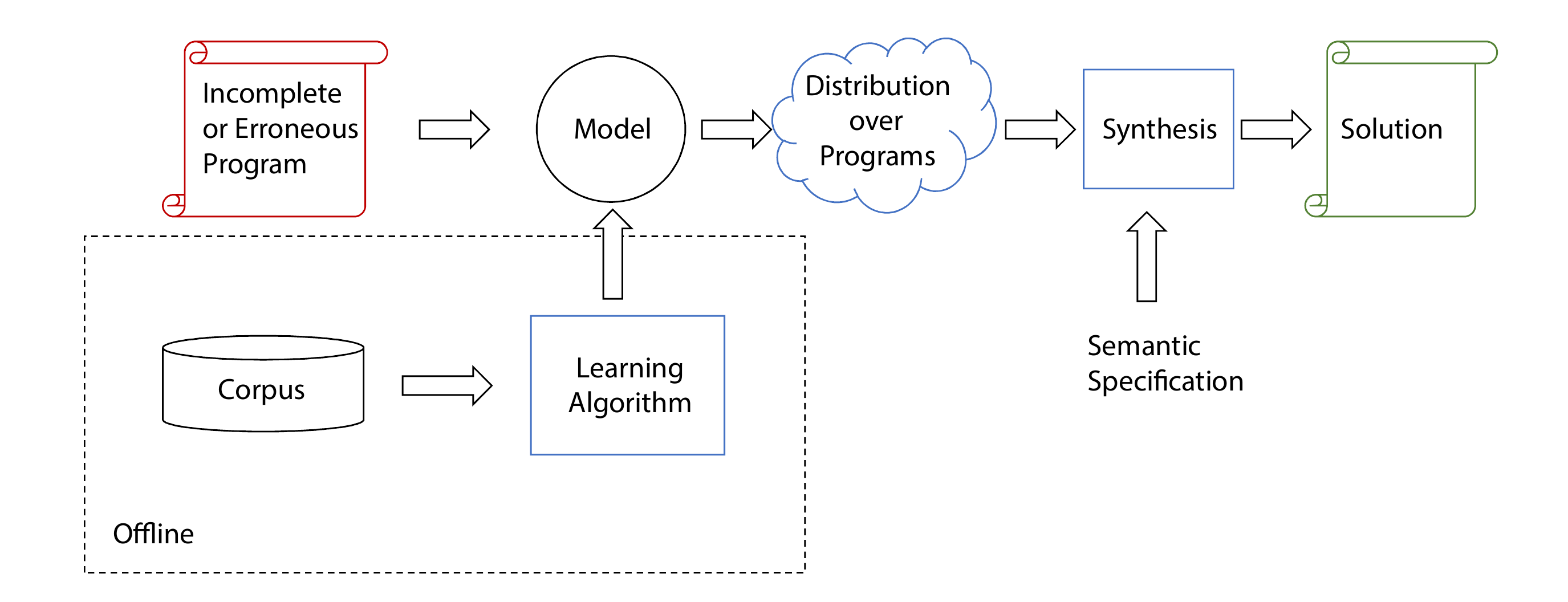}
\caption{Data Driven Synthesis Framework}
\figlabel{ddsfigure}
\end{figure}

\subsection{Data Driven Corrections for MOOCs}
We have developed a DDS-based system called $sk\_p$ that can correct small programming assignments in Python. $sk\_p$ innovates on the general DDS paradigm in three important respects, all suited to the characteristics of our domain. First, $sk\_p$ constructs models that are \emph{purely syntactic}; the model treats a program statement as a list of tokens and assumes no further program structure or semantics, aside from a distinction between whether a token is a variable name or not.
This is in contrast to prior approaches to DDS which rely heavily on features derived from program analysis and which learn from a more structured representation of programs. 

Secondly, we use a modified seq2seq neural network~\cite{seq2seq}, which learns the syntactic structures of program statements and is able to produce valid statements for a candidate program. The neural networks are trained on a corpus of \emph{correct} programs, where the correctness is established via the same test suite used to validate candidate solutions. The neural-network model is generative, which implies that we can easily use it to sample from the space of possible fixes; This is in contrast to the models used by prior repair work where the model was discriminative, and therefore the synthesis algorithm had to explicitly enumerate a large space of possible corrections to find the one with the highest probability~\cite{fan}. 


A third surprising aspect of our solution is that the models are very local: At each correction site, the model only uses one statement before and after the site as context to generate a distribution of corrections, ignoring the rest of the program. This model is called a skipgram, a popular model used in NLP in the task of word embedding~\cite{word2vec}. In essence, our method learns short code fragments that appear frequently in correct solutions and identifies fragments in incorrect submissions that look similar. We show that this very local model is actually accurate enough that the synthesis component of DDS can quickly find a correct solution with a simple enumerate-and-check strategy. 

\subsection{Results}

We evaluate $sk\_p$ on 7 different Python programming assignments from an early version of 6.00x in MITx. The training sets range in size from 315 to 9078 problems, and resulting models are tested on a separate set of incorrect programs of which $sk\_p$ can correct 29\%. The details of the experiments are explained in \ref{experiments}, but overall, our empirical evaluation allows us to make the following observations: 

\paragraph{$sk\_p$ is competitive with Autograder:} Of the 7 benchmarks assignments, autograder \cite{rishabh} provides correction models for 3 assignments which can generate good quality feedback in real-time (under 5 seconds per submission) at an average accuracy of 30\%. $sk\_p$, which has an average runtime of 5.6 seconds, outperforms autograder marginally with an average accuracy of 35\% on these 3 assignments. This is surprising given the fact that our system does not rely on the instructor to provide a correction model, and its only knowledge of the python semantics comes from its ability to run the python interpreter off-the-shelf.

\paragraph{Syntactic errors matter:} On average, 18\% of $sk\_p$'s corrections are fixing syntactic errors; On certain benchmarks, syntactic errors account for 40\% of the fixes. These experiments highlight the importance of handling programs with syntactic errors which do not parse.

\paragraph{Efficacy of Neural Network:} We evaluate our neural network model on the task of fragment learning by considering an alternative, exhaustive model that explicitly memorizes all the program fragments during training. We find that the neural network out-performs the exhaustive model when there is a sufficient number of training programs relative to the total number of fragments that needs to be learned. The neural network's average accuracy of 29\% comes close to the average accuracy of 35\% of the exhaustive model.

\subsection{Contributions}

The paper makes the following contributions:
\begin{itemize}
\item \textbf{Correction by Fragment Completion:} We validate a hypothesis that using fragment completion as a mechanism for correction, recalling similar fragments from correct programs, works well in the context of MOOCs.
\item \textbf{Purely Syntactic Learning:} The fragment completion model using neural networks is purely syntactic: it treats a program statement as a \emph{sequence of tokens}, with the candidate missing statement generated verbatim one token at a time. A direct consequence of this syntactic learning is the ability to fix syntactic errors, without requiring the seed program to parse.

\item \textbf{Learned Correction Model:} Compared to prior work where a different, manual correction model is required for each assignment, the specifics of how to complete a fragment are learned from data.

\item \textbf{Simple Synthesis Procedure:} The fragment completion model using neural networks generates program statements that parse 	with high probability; these statements are used directly to form a candidate program without further constraint solving. As a result our synthesis procedure does not need to perform analysis on the candidate programs, and is a simple enumerate and check framework using the test suite.
\end{itemize}
The rest of the paper elaborates on the details of our technique. 

\section{Overview}

Consider the programming assignment of writing a function to evaluate an uni-variate polynomial, represented as a list of coefficients ($poly$), at a point $x$. Below is a student solution which is incorrect:

\begin{lstlisting}
def evaluatePoly(poly, x):
  a = 0
  f = 0.0
  for a in range(0,len(poly) $-$ 1):
    f = poly[a]*x**a+f
    a += 1
  return f
\end{lstlisting}

This code would have been correct if the for-loop is allowed to iterate to the full length of the input $len(poly)$. However, $sk\_p$ was able to correct this program differently as follows:

\begin{lstlisting}
def evaluatePoly(poly, x): 
  a = 0 
  f = 0.0 
  while a < len(poly): 
    f = poly[a]*x**a+f 
    a += 1 
  return f 
\end{lstlisting}

We see $sk\_p$ replaced the for-loop with a while-loop. While removing the $-1$ at the end of the for loop, a small local modification, would also produce a semantically correct program, the correction suggested by $sk\_p$ is both semantically correct and more natural. We now give a high level overview of our correction algorithm, starting from the incorrect program and ending at the correct program.

\subsection*{Renaming Variables} 
In $sk\_p$, a program statement is represented syntactically as a sequence of tokens. A key assumption with this representation is the existence of a \emph{finite sized vocabulary}: when modeling a sentence as a sequence of words in NLP, a dictionary is customarily used to bound the total number of words. We bound the total number of tokens by renaming variable names in a primitive manner: keywords such as ``if'', ``for'', common function names and method names such as ``len'', ``range'', along with the arithmetic operators are specified to be excluded from renaming. Any unspecified tokens are renamed from $x_0$ up to $x_K$. For a given assignment, the upper bound for $K$ across all submissions is typically small. Here is the resulting program from renaming the variables:

\begin{lstlisting}
_start_ 
  x2 = 0 
  x3 = 0.0 
  for x2 in range ( 0 , len ( x0 ) $-$ 1 ) : 
    x3 = x0 [ x2 ] * x1 ** x2 + x3 
    x2 += 1 
  return x3
_end_
\end{lstlisting}

Note that we only represent the body of the function definition, and an artificial start and end statement are padded around the statements, which will help in forming the program fragments later.

\subsection*{Forming Partial Fragments} 
In $sk\_p$, we consider the program fragments of 3 consecutive statements. A fragment is formed for each of the original statement in the program, consisting of its previous statement, itself, and the next statement. In the actual implementation, we also consider other form of fragments which allow the algorithm to insert and remove statements. Here are the first three fragments of our example problem:

Fragment 1:

\begin{tikzpicture}[node distance = 0.0in]
\node(a)[anchor=west]{
\begin{lstlisting}
_start_ 
  x2 = 0 
  x3 = 0.0 
\end{lstlisting}
};
\end{tikzpicture}

Fragment 2:

\begin{tikzpicture}[node distance = 0.0in]
\node(a)[anchor=west]{
\begin{lstlisting}
  x2 = 0 
  x3 = 0.0 
  for x2 in range ( 0 , len ( x0 ) $-$ 1 ) : 
\end{lstlisting}
};
\end{tikzpicture}

Fragment 3:

\begin{tikzpicture}[node distance = 0.0in]
\node(a)[anchor=west]{
\begin{lstlisting}
  x3 = 0.0 
  for x2 in range ( 0 , len ( x0 ) $-$ 1 ) : 
    x3 = x0 [ x2 ] * x1 ** x2 + x3 
\end{lstlisting}
};
\end{tikzpicture}

For these fragments, the original program statement in the middle is removed forming \emph{partial fragments}, consisting of the two surrounding statements and a ``hole'' for the missing statement:

Partial Fragment 1:

\begin{tikzpicture}[node distance = 0.0in]
\node(a)[anchor=west]{\ttfamily{_start_}};
\node(b)[below=of a.south west, anchor=north west, draw, minimum width=1in]{
\begin{lstlisting}

\end{lstlisting}
};
\node(c)[below=of b.south west, anchor=north west]{\ttfamily{  x3 = 0.0}};
\end{tikzpicture}

Partial Fragment 2:

\begin{tikzpicture}[node distance = 0.0in]
\node(a)[anchor=west]{\ttfamily{  x2 = 0}};
\node(b)[below=of a.south west, anchor=north west, draw, minimum width=1in]{
\begin{lstlisting}

\end{lstlisting}
};
\node(c)[below=of b.south west, anchor=north west]{\ttfamily{  for x2 in range ( 0 , len ( x0 ) - 1 ) :}};
\end{tikzpicture}

Partial Fragment 3:

\begin{tikzpicture}[node distance = 0.0in]
\node(a)[anchor=west]{\ttfamily{  x3 = 0.0}};
\node(b)[below=of a.south west, anchor=north west, draw, minimum width=1in]{
\begin{lstlisting}

\end{lstlisting}
};
\node(c)[below=of b.south west, anchor=north west]{\ttfamily{    x3 = x0 [ x2 ] * x1 ** x2 + x3}};
\end{tikzpicture}

In order to generate the distribution of candidate programs, $sk\_p$ will pass each of these fragments to the \emph{statement prediction model} which will generate a list of likely candidate statements that should fill the hole, possibly forming program fragments that resembles that of a correct program. 

\subsection*{Predicting Statements from Partial Fragments} 
The statement prediction model is tasked with generating candidate missing statements, using the partial fragment as context. 
We briefly describe how the model is trained and explain how it works on a high level.

\paragraph{Training:} Our statement prediction model is first trained on a corpus of fragments from correct programs. Each correct fragment is converted to an input-output training pair: The partial fragment (with a hole) is the input, and the missing statement is the output. For instance, here is one of the training input-output pair, derived from a similar correct fragment:

Example Training Input:

\begin{tikzpicture}[node distance = 0.0in]
\node(a)[anchor=west]{\ttfamily{else:}};
\node(b)[below=of a.south west, anchor=north west, draw, minimum width=1in]{
\begin{lstlisting}

\end{lstlisting}
};
\node(c)[below=of b.south west, anchor=north west]{\ttfamily{    x2 += x0[x3] * (x1 ** x3)}};
\end{tikzpicture}

Example Training Output:

\begin{tikzpicture}[node distance = 0.0in]
\node(a)[anchor=west]{\ttfamily{}};
\node(b)[below=of a.south west, anchor=north west, draw, minimum width=1in]{
\begin{lstlisting}
  while x3 < len ( x0 ) :
\end{lstlisting}
};
\node(c)[below=of b.south west, anchor=north west]{\ttfamily{}};
\end{tikzpicture}

\paragraph{Statement Prediction Model:} Our model is implemented using a neural network, using architecture inspired by the seq2seq \cite{seq2seq} network and the skip-thought network \cite{skip-thought}. The seq2seq network has been traditionally used in machine translation. The seq2seq network consists of an \emph{encoder} and a \emph{decoder}: the encoder reads the input sequence of words (say in English) one word at a time, and updates an internal state each time. When the encoder finishes reading the input sequence, its internal state represents a high level summary of the English sentence. This state is then passed into a decoder, which generates words (say in Spanish) one word at a time via sampling, effectively translating the English sentence into Spanish. 

Our statement prediction model is almost identical to the seq2seq architecture, except instead of one encoders we use two different encoders, one for the preceding statement in the partial fragment, and one for the following statement. The two encoders summarize each statement independently, and their summaries are joined together and passed to the decoder to generate the candidate missing statement via sampling. A particular instance of this sampling is shown in Figure~\ref{fig:encdec}. In the actual implementation, we use beamsearch, a deterministic algorithm that is guaranteed to return high probability statements instead of actual sampling.

\paragraph{Candidate Statement Generation:} For Partial Fragment 3, our model produces the following candidate statements, along with their probabilities conditioned on the partial fragment.

Input (Partial Fragment 3):

\begin{tikzpicture}[node distance = 0.0in]
\node(a)[anchor=west]{\ttfamily{  x3 = 0.0}};
\node(b)[below=of a.south west, anchor=north west, draw, minimum width=1in]{
\begin{lstlisting}

\end{lstlisting}
};
\node(c)[below=of b.south west, anchor=north west]{\ttfamily{    x3 = x0 [ x2 ] * x1 ** x2 + x3}};
\end{tikzpicture}

Top 3 output candidate statements with probabilities:

\begin{tikzpicture}[node distance = 0.0in]
\node(a)[anchor=west]{\ttfamily{}};
\node(b)[below=of a.south west, anchor=north west, draw, minimum width=1in]{
\begin{lstlisting}
0.141, while x2 < len ( x0 ):
0.007, for x4 in range ( len ( x0 ) ) :
0.0008,for x4 in range ( 0 ) :
\end{lstlisting}
};
\end{tikzpicture}

$sk\_p$ computes a distribution for \emph{every} partial fragment in the original program. Here we show the candidate statements, with probabilities, on the first two fragments:

Candidate Statements for Partial Fragment 1:

\begin{tikzpicture}[node distance = 0.0in]
\node(a)[anchor=west]{\ttfamily{}};
\node(b)[below=of a.south west, anchor=north west, draw, minimum width=1in]{
\begin{lstlisting}
0.321, x2 = 0
0.009, x2 = len ( x0 [ 0 ]
0.008, x2 = 0.0
\end{lstlisting}
};
\end{tikzpicture}

Candidate Statements for Partial Fragment 2:

\begin{tikzpicture}[node distance = 0.0in]
\node(a)[anchor=west]{\ttfamily{}};
\node(b)[below=of a.south west, anchor=north west, draw, minimum width=1in]{
\begin{lstlisting}
0.00013, x3 = 0
1.77e$-$6, x3 = 0.0
8.55e$-$8, x3 = 1
\end{lstlisting}
};
\end{tikzpicture}

Note that the neural network is not guaranteed to generate syntactically correct fragments, as illustrated by the results from Partial Fragment 1.

\subsection*{Finding a Candidate Program} 

The model produces a distribution of corrections for every statement in the program. Note, however, that in the case of the first and second statements, the highest probability statements leave the statement unchanged or almost unchanged, whereas for the third statement, the highest probability candidate is the replacement necessary to fix the program, 
although that will not always be the case. In general, the distribution over the space of all possible combinations of corrections needs to searched explicitly.

\paragraph{Space of Candidate Programs} To form a candidate program, $sk\_p$ considers the set of candidate programs derived by applying \emph{simultaneous replacement} on every line of the original program, choosing a statement from the set of candidate statements of its partial fragment to replace it.

For instance, if we replace \emph{all} the original statements by choosing the 2nd candidate statement, the resulting candidate program would have these first 3 lines:

\begin{tikzpicture}[node distance = 0.0in]
\node(a)[anchor=west]{
\begin{lstlisting}
  x2 = len ( x0 [ 0 ]
  x3 = 0.0
  for x4 in range ( len ( x0 ) ) :
  ...
\end{lstlisting}
};
\end{tikzpicture}

The process of simultaneous replacement is generalized in our work so that we can generate candidate programs that also have insertions and deletion of statements by considering other form of fragments. In our work we also consider the original statement as one of the candidates (even if it did not appear in the generated distribution) so $sk\_p$ always has the option of not replacing the original statement. This is useful when an unfamiliar fragment is given to the statement prediction model, and the model cannot give high quality candidate statements. This is explained in more detail in Section 5. 

\paragraph{Distribution of Candidate Programs} We define the probability of a particular candidate program as the product of the probabilities of its chosen candidate statements. The search procedure uses the top-k candidates (generated by beam search) from the statement prediction model, and sort them into a priority queue based on their probabilities. Each enumerated candidate program is checked for correctness against the spec, and the first correct program (the one with the highest probability and also correct) is returned. For our example, it is this one:

\begin{tikzpicture}[node distance = 0.0in]
\node(a)[anchor=west]{
\begin{lstlisting}
  x2 = 0 
  x3 = 0.0 
  while x2 < len ( x0 ) : 
    x3 = x0 [ x2 ] * x1 ** x2 + x3 
    x2 += 1 
  return x3 
\end{lstlisting}
};
\end{tikzpicture}

This correct program is the 337th program to be enumerated. Once a correct program is found, the variable names are returned to their original names, and the program is given back to the student as feedback.

\section{Correction Model}

Our approach fixes an incorrect program by altering it via replacements, insertions, and deletions. These changes are applied on a statement level: An entire statement is inserted or replaced. To decide how these changes are applied, we use a method which we call \emph{Fragment Completion}. For each statement in the incorrect program, we consider the program fragments consisting of itself and its previous and next statements. We then ask whether this program fragment can be made to more resemble a program fragment from a known correct program. This is done by removing the original statement, forming a partial fragment consisting of just the surrounding statements, then completing the partial fragment with other statements. 

\subsection{Skipgram Models}

Our work is heavily inspired by Skipgram models, which have been widely used in natural language processing~\cite{word2vec,glove} to learn lexical semantics in terms of co-occurrence patterns. Consider the sentence ``I like to write computer programs with an editor.''. The word \emph{programs} has other words such as \emph{computer} and \emph{editor} occurring around it in the sentence, which are correlated. Skip-gram models utilize these correlations to learn vector representations for the words such that semantically similar words have comparable representations. In fact, if we were to hide the word \emph{program} away, one can still conceivably recover this word back by looking at its surrounding words such as \emph{computer} and \emph{editor}.

Recent work has extended the idea of the basic Skipgram model to the sentence level~\cite{skip-thought} where instead of a sequence of words, the correlations between a set of \emph{sentences} are considered. 

In our work, we explore the following question: Instead of words and sentences, what if statements in a code fragment are correlated in a similar way? The programming equivalent to a skipgram, which is made of words and sentences, is that of a \emph{Partial Program Fragment}, consisting of a pair of program statements with a hole in the middle, which can be completed with an \emph{Missing Statement}. We show this simple model can be adapted for program correction that is able to correct a wide varieties of mistakes.

\subsection{Statement Transformations by Fragment Completion}

Our method corrects an incorrect program by applying to it a series of statement transformations. A statement transformation alters the incorrect program $\mathcal{X}$ by either replacing an existing statement, inserting a new statement, or deleting a statement. These transformations are not applied in a vacuum. Specifically, each transformation also has a context of partial fragments in which to inform its decision. Formally, a statement transformation is a pair: The first element is a partial fragment, the two surrounding statements where the transformation is applied; The second element is a candidate statement used to perform the transformation by completing the partial fragment. 

Let the incorrect program be denoted as $\mathcal{X} = X_0 \dots X_{n+1}$ where $X_i$ is the i-th statement of the incorrect program, and padded with an artificial ``begin'' statement $X_0$ and an ``end'' statement $X_{n+1}$. We now formally define 3 kinds of statement transformations:

\begin{itemize}
\item \emph{Replacement} $R_i = ((X_{i-1},X_{i+1}), Y_i)$: The statement $X_i$ is to be replaced by the candidate statement $Y_i$. The partial fragment context for replacing statement $X_i$ is the surrounding statements $X_{i-1}$ and $X_{i+1}$.
\item \emph{Insertion} $I_i = ((X_i, X_{i+1}),Y_{i,i+1}$: A new candidate statement $Y_{i,i+1}$ is to be inserted between the statements $X_i$ and $X_{i+1}$, $X_i$ and $X_{i+1}$ also serve as the context for insertion.
\item \emph{Deletion} $D_i$: The statement $X_i$ should be removed. This is realized using the replacement transformation $R_i = ((X_{i-1},X_{i+1}), \epsilon)$, where instead of a candidate statement $Y_i$ we replace $X_i$ by the empty statement $\epsilon$.
\end{itemize}

Note we can express the null transformation under this scheme:
\begin{itemize}
\item $R_i = ((X_{i-1},X_{i+1}), X_i)$: This will replace $X_i$ with $X_i$ itself, causing no change.
\item $I_i = ((X_i, X_{i+1}), \epsilon)$: An empty statement is inserted between between statements $X_i$ and $X_{i+1}$.
\end{itemize}
The null transformation will be useful in section 5 where we address the issue with unseen skipgrams.

\begin{figure}[!ht]
  \centering
    \includegraphics[width=0.5\textwidth]{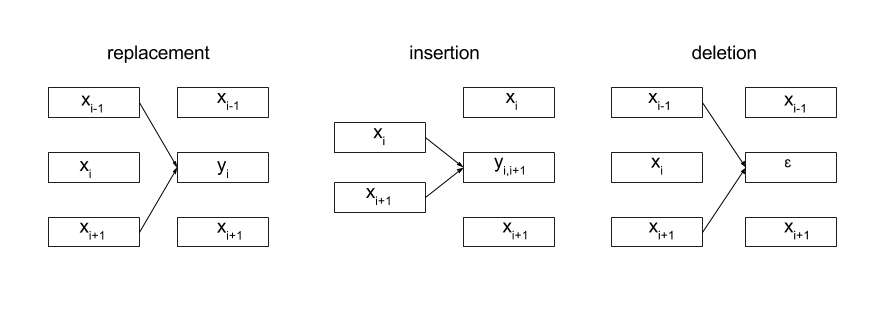}
      \caption{Statement Transformations}
  \label{fig:stmt_xform}
\end{figure}

The three kinds of statement transformations are shown in Figure \ref{fig:stmt_xform}. For each transformations, the partial fragment context statements are connected to the tails of the arrows, and the candidate statement that completes the fragment is pointed to by the heads of the arrows. The direction of the arrows indicates that we are predicting the candidate statement from its fragment context.

\section{Statement Prediction Model}
\label{subs:model}

We notice that although there are 2 different kinds of corrections: replacement and insertion (deletion is modeled as replacement by the empty statement), they all share the same fragment completion structure: A candidate statement is used to complete its partial fragment context. This suggest a single model can be trained for both replacement and insertion tasks instead of two separate models. We formalize the prediction task as follows:

Given a pair of program statements $X, X'$ as context, predict a list of likely candidate statements $Y^1 \dots Y^K$ that can exist between $X$ and $X'$ (note: the candidate statements can be the empty statement $\epsilon$ as well), along with their probabilities $Pr(Y^j | X, X')$. We ask the prediction model for a list of candidates instead of a single candidate because given a skipgram context, there might be multiple ways of completing it in a correct program. Therefore, a conditional probability distribution $Pr(Y | X, X')$ is constructed and the top $k$ candidates are chosen from this distribution.

\subsection{Generating Training Fragments}
To obtain the distribution $Pr(Y | X, X')$, we first need to train the model on a corpus of correct program fragments. 

Our dataset is not particularly large for some of the benchmarks. For comparison in \cite{giga}, gigabytes of natural language corpus being read to train a language model. As a result we might have a problem of data sparsity due to our relatively small dataset, and our model will have a hard time finding patterns in the (irregular) training data.

To resolve this, we apply a rudimentary regularity filter to the training programs, using a correct program for training only if:
\begin{itemize}
\item the number of lines in the solution is smaller than a bound seq\_n
\item the maximum number of tokens within a statement is smaller than a bound seq\_l
\item the tokens a program use is within the set of commonly used tokens freq\_toks
\end{itemize}

The bound seq\_n is computed separately for each benchmark by first collecting the number of lines used in all the submissions, and taking the bound such that 97\% of the programs have line number less than it. The bound seq\_l is computed similarly, except by first collecting the maximum length of any statement of a program  to a benchmark. The set of commonly used tokens is collected by counting all the token usages across the programs in a benchmark, and taking the top 99.9\% of the most frequently used tokens. For our benchmarks, the regularized data is about 90\% of the unregularized data.

From the regularized training data, we set up the training for the skipgram language model as follows:

Given a correct program $X$ of $n$ statements $X_1 \dots X_n$, we first pad the program statements with two artificial statements $X0$ and $X_{n+1}$ on front and back forming $n+2$ statements $X_0, X_1 \dots X_n, X_{n+1}$. Then, for every 2 consecutive statements in the padded statements, we generate the training data:

\[
 (X_i, X_{i+1}) \Rightarrow \epsilon ~\forall i \in 0 \dots n
\]

This training data express the following correction strategy: Given the partial fragment that resembles two consecutive statements from a correct program, $X_i, X_{i+1}$, no statement should exist between them.

Also, for every original statement $X_i \in \{X_1 \dots X_n\}$ we generate a training data:
\[
 (X_{i-1}, X{i+1}) \Rightarrow X_i ~\forall i \in 1 \dots n
\]

This training data pair express a different correction strategy: Given a partial fragment that resembles the surrounding statements of $X_i$ from a correct program, the statement $X_i$ should exist between them. 

\begin{figure}[!ht]
  \centering
    \includegraphics[width=0.5\textwidth]{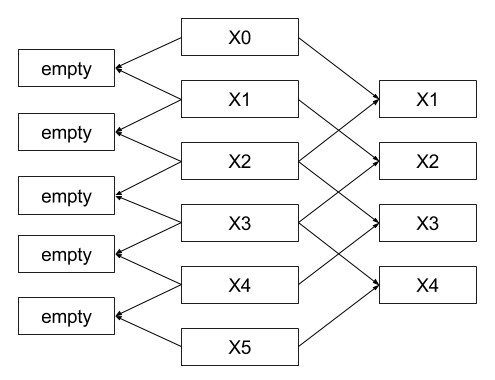}
  \caption{Generating Training Input Output}
  \label{fig:datagen}
\end{figure}

We shows how to generate data from a correct program in Figure~\ref{fig:datagen}. Here, each pair of arrows represents a pair of training data from the input partial fragment to the output missing statement.

\subsection{Neural Network Model}

We now explain the implementation of the statement prediction model. In this work, we propose an encoder-decoder model for statement prediction using recurrent neural networks (RNN). Rather than storing the program fragments inputs and the candidate statement outputs verbatim, this model is capable of reading the context statements as inputs, and generate candidate statements as outputs. 

\subsubsection*{Tokenization and Variable Renaming}
To use RNN in the task of statement prediction, we think of each statement as a sequence of atomic tokens such as variables, keywords, indentations, and arithmetic operators. One key concern is the issue of unbounded number of tokens: For our approach to work, the total number of tokens need to be bounded, yet students are free to invent an arbitrary number of variable and function names. To solve this issue, we rename the variables in the student submissions, and since the total number of variables are typically bounded in the MOOCs setting, the total number of tokens after renaming are bounded. We do not rename functions since the student implement programs that consist of a single function without helpers.

Our algorithm tokenizes the entire student solution, then for each named token in the solution, attempts to rename it to $x_i$ where $i$ is the ith unique named token seen by the algorithm. To distinguish a variable name such as ``myTup'' from a useful function name such as ``range'', we apply the following strategy: First, a token statistic is collected across all correct student submissions, recording all named tokens along with the frequencies these tokens are being used in the submissions. Second, we construct a list of ``forbidden'' names: a list of names that should not be renamed. In the beginning, the forbidden list is empty, which would cause useful functions such as $range$ to be renamed $x_i$, causing all correct programs to become incorrect. The list of forbidden words is gradually grown by introducing the most frequent tokens from the token statistic, thus, useful functions such as $range$ that is used in every submission, along with common variable names such as $x$ are no longer being renamed. This growth continues until the number of correct programs reaches 98\% of the original number of correct programs, with obscure library calls still being re-named at the expense of correctness. Then, this forbidden list is reduced by attempting to remove each token from the list: If a common variable $x$ is removed, the number of correct programs would not change, but if an actual function $range$ is removed, the number of correct programs would decrease. By the end, we would have obtained a list of useful function names which should not be renamed.

Once tokenized, one can rewrite the skipgram statements and the candidate statement as a sequence of tokens as follows:

\begin{align*}
\begin{split}
	X &= x_1, x_2, \dots x_N \\
    X' &= x'_1, x'_2, \dots x'_M \\
    Y &= y_1, y_2, \dots y_R \\
\end{split}
\end{align*}

\subsubsection*{Recurrent Neural Network and LSTM}
We now briefly describe RNN, which are widely used to model sequential data. Conceptually, an RNN captures sequential computation by using RNN cells, which is a parametrized update function that processes an input at each timestep. The RNN cell takes in an input $x_{t}$ (the current data at iteration $t$) and a previous hidden state $h_{t-1}$ as arguments, and produces two outputs: the current hidden state $h_t$, and a \emph{distribution} of possible values for the output $y_t$, $Pr(y_t | h_{t-1}, x_{t})$. For our case, each input and output for the RNN has as many possible values as there are distinct number of tokens. Figure REF depicts this high level view. We use the generic letter $\theta$ to denote all the learnt parameters of the RNN cell.

In this work, we employ LSTM ~\cite{lstm}, which is a particular implementation of the RNN cell that works well in remembering long term dependencies. In an LSTM, the hidden state $h_t$ is comprised of 2 parts, the hidden cell state $c_{t}$ and the output $y_{t}$. The rationale behind this is that the hidden cell state $c_t$ is now used to primarily remember long term dependencies, while the output $y_{t}$ is used as a proxy for short-term dependencies. The input, output, hidden state, and parameters are encoded as continuous valued vectors. In particular, the input and output vectors of length $N_{tk}$, the number of possible distinct tokens, where the value at the $i^{th}$ index denotes the probabilities of the input(or output) takes on the value of the $i^{th}$ token. and the LSTM cell as a function is expressed as a set of update equations:

\begin{align}
\label{eqn:eqlabel}
\begin{split}
	i_t &= \sigma (U^{(i)} x_t + V^{(i)} y_{t-1} + b^{(i)}), \\
	f_t &= \sigma (U^{(f)} x_t + V^{(f)} y_{t-1} + b^{(f)}), \\
	o_t &= \sigma (U^{(o)} x_t + V^{(o)} y_{t-1} + b^{(o)}) \\
	z_t &= \tanh ( U^{(z)} x_t + V^{(z)} y_{t-1} + b^{(z)})	\\
	c_t &= i_t \odot z_t + f_t \odot c_{t-1} \\
	y_t &= o_t \odot \tanh(c_t)
\end{split}
\end{align}

Here, $\sigma$ represents the sigmoid function and $\odot$ is elementwise multiplication. $U^{(i)}, U^{(f)}, U^{(o)}, U^{(z)}$ and their $V$ and $b$ counterparts are parameters (expressed as matrices) learnt by the model. To represent that an input $x_t$ is the $i^{th}$ token, it is modeled as a 1-hot vector, having a value of $1$ at the $i^{th}$ index and $0$ everywhere else. Similarly, the vector $y_t$ can be normalized (using a soft-max function) and the value at the $i^{th}$ position denotes the probability that $y_t$ being the $i^{th}$ token.

For clarity, we will use the high level RNN formulation where we denote the hidden state by $h_t$.

\subsubsection*{Encoder Decoder Model}
\begin{figure}[!ht]
  \centering
    \includegraphics[width=0.5\textwidth]{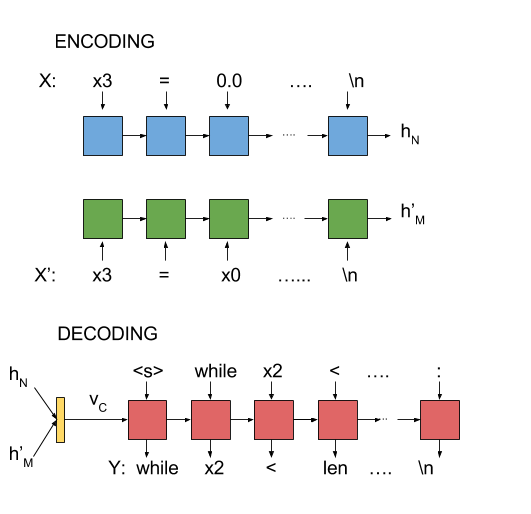}
  \caption{Encoder Decoder Model}
    \label{fig:encdec}
\end{figure}

We use RNN in two forms: (1) as an \emph{encoder} to output a vector $v_C$ representing a summary for the context statements, and (2) as a \emph{decoder} to generate a candidate statement $Y$ given the context vector $v_C$. Figure \ref{fig:encdec} shows an overview of our encoder decoder architecture generating a candidate statement from its skipgram context.

To encode the two skipgram context statements $X, X'$, we use two different encoders colored blue and green, one for each statement. For the encoding task, we only care about the hidden states (horizontal arrows in the encoding network) , which contains a summary of all the  prefix of the input sequence at each step. The last hidden states are labeled $h_N$ and $h'_M$, they are vectors representing the overall summary of the input $X$ and $X'$ respectively. These two vectors are concatenated (forming a single, longer vector) and passed through a linear layer (a matrix of learnt parameters，yellow in the figure) to obtain the context vector $v_C$. The outputs of these RNNs are not used for encoding, and are not shown in the figure.

Now, from $v_C$, we \emph{generate} an output statement $Y$ by using a decoder RNN, colored red. As the context vector $v_C$ serves as a summary for the context $X X'$, we can rewrite $Pr(Y | X X') = Pr(Y | v_C)$. We will first explain the generation of $Y$ as a random sampling process from the distribution $Pr(Y | v_C)$, then briefly describe beam-search, which is a way of reliably generating an approximated top-k candidates from the distribution without random sampling.

To obtain a sample from the distribution $Pr(Y | v_C)$, we first rewrite $Y$ as a sequence of tokens, then factor it using conditional probabilities:

\begin{align}
\begin{split}
	& Pr(Y | v_C) \\ 
     &= Pr(y_1, y_2 \dots y_R | v_C)\\
     &= Pr(y_1 | v_C) Pr(y_2 | v_C, y_1) \dots \\
     &  ~~~~~Pr(y_R | v_C, y_1 \dots y_{R-1})\\
\end{split}
\end{align}

We now show how the decoder RNN computes each of the conditional probabilities in the product of the last equation. At the first step, the RNN cell takes in a fixed artificial start symbol $\langle s \rangle$ as input, along with the context vector $v_C$ (as the first hidden state $v_C = h_0$) to produce the first hidden state $h_1$, and the conditional distribution for the first output token $Pr(y_1 | v_C)$. We sample from this distribution to obtain the first output token. This output token is fed back into the RNN cell at the second step, along with the hidden state $h_1$ to obtain the hidden state $h_2$ and the conditional distribution $Pr(y_2 | v_C, y_1) = Pr(y_2 | h_1, y_1)$, and again we sample this distribution for the second token. This process continues, and at each step $t$ we sample from the conditional distribution $Pr(y_t | v_C, y_1 \dots y_{t-1}) = Pr(y_t | h_{t-1}, y_{t-1})$ for the $t^{th}$ token, where the hidden state $h_{t-1}$ is used to capture the dependency on all previously generated tokens.

Hence, we have the likelihood of the entire sequence generated by the decoder (via sampling) as follows:
\begin{dmath*}
Pr(y_1 \dots y_R | v_C) = \prod_t Pr(y_t | h_{t-1}, y_{t-1})
\end{dmath*}

We have now described how to use the encoder-decoder architecture with RNNs to sample from the distribution $Pr(Y | X X')$. Conceivably, one can repeat the sampling process many times and take the top-k candidates for the prediction task, but it may require many samples and be expensive.

A better alternative to sampling is to use a \emph{Beam Search} \cite{beam}, which we will briefly describe. Rather than building a single candidate statement one token at a time, in beam search, we keep track of the top-k candidate prefixes. We deterministically choose the top-k tokens from the distribution $Pr(y_t | h_{t-1}, y_{t-1})$ and store all possible ways of growing the top-k prefixes by appending these tokens. This would cause an explosion of number of candidates to be stored, thus we prune the candidates according to the prefix probability $Pr(y_1 \dots y_t | v_C)$ to keep the total number of candidate prefixes under $k$.

In our example, the top 3 candidates decoded by our beamsearch are as follows:

\begin{tikzpicture}[node distance = 0.0in]
\node(a)[anchor=west]{\ttfamily{}};
\node(b)[below=of a.south west, anchor=north west, draw, minimum width=1in]{
\begin{lstlisting}
0.141, while x2 < len ( x0 ):
0.007, for x4 in range ( len ( x0 ) ) :
0.0008,for x4 in range ( 0 ) :
\end{lstlisting}
};
\end{tikzpicture}

\paragraph{Implementing the Statement Prediction Model:} The statement prediction model is implemented using the TensorFlow \cite{flow} framework. In particular, the two encoder cells and the decoder cell are implemented as a 2-layer stacked LSTM with 50 hidden units each. The network is trained using batched gradient descent with a batch of size 50, and optimized using the RMSProp optimizer. The training is done over 50 epochs, at each epoch, we measure the cross entropy loss on the validation set, with the lowest cross entropy of the 50 epochs stored.

\section{Generating Candidate Programs}

So far we described a statement prediction model: Given a pair of statements $X X'$ as context, it will generate a list of top-k candidates $Y^1 \dots Y^K$ that can exist between $X$ and $X'$. To use this model for correction, however, requires another piece of information: Where should the correction happen?

One can train yet another model for the error localization task. Given an incorrect program, this model will predict the locations to perform the statement replacements and insertions. Training this model would require a pair of of programs $\mathcal{X}, \mathcal{Y}$ such that $\mathcal{Y}$ is a correction for $\mathcal{X}$. In this work, we opt for a simpler approach by using the statement prediction probabilities to perform the localization task implicitly: Given an incorrect program $\mathcal{X} = X_0 \dots X_{n+1}$ (with padded artificial statements $X_0$ and $X_{n+1}$), we put all the statements $X_1 \dots X_n$ up for replacement using our statement prediction model. The rationale is that a correct statement $X_i$ is more likely to be present between the skipgram $X_{i-1}$ and $X_{i+1}$ than an incorrect statement. Therefore, if we use the statement prediction model to replace $X_i$, with high probability our prediction model would \emph{regenerate} $X_i$ back, which is equivalent to identifying that $X_i$ does not need to be corrected. On the otherhand, suppose a statement $X_j$ is incorrect, then with high probability the statement prediction model would produce a different statement $X_j'$ in its place to complete the skipgram $X_{j-1}, X_{j+1}$, effectively identifying that this statement needs to be replaced, and providing the candidates. This implicit localization is also performed for the insertion task by considering insertion between all pairs of statements from $\mathcal{X}$. If an insertion is not needed, we leave it up to our prediction model to predict the empty statement $\epsilon$ with a high probability. 

Given an incorrect program $\mathcal{X}$, we obtain a candidate program $\mathcal{Y}$ by applying a single statement replacement on \emph{each} of the existing statements of $\mathcal{X}$, and applying a single insertion between \emph{each} pairs of consecutive statements in $\mathcal{X}$ as well. To decide which candidate statements should be chosen for these replacements and insertions, we perform a search over the space of all possible candidate programs, $\bar{\mathcal{Y}}$, which is defined as follows: 

\begin{align}
\begin{split}
	& \bar{\mathcal{Y}}  = \overbar{Y_{0,1}} \times \overbar{Y_{1}} \times \overbar{Y_{1,2}} \times \overbar{Y_{2}} \times \dots \times \overbar{Y_{n}} \times \overbar{Y_{n-1,n+1}}\\ 
     & \overbar{Y_{i}} = [Y_{i}^{1} \dots Y_{i}^{K}]\\
     & \overbar{Y_{i,i+1}} = [Y_{i,i+1}^{1} \dots Y_{i,i+1}^{K}]\\
\end{split}
\end{align}

\begin{figure}[!ht]
  \centering
    \includegraphics[width=0.5\textwidth]{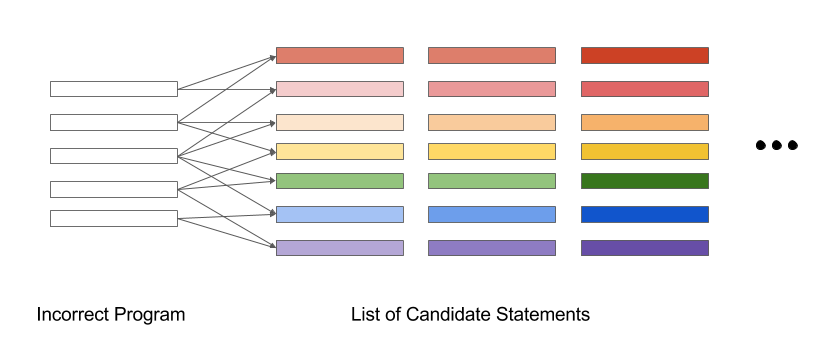}
  \caption{Space of Candidate Programs}        \label{fig:sspace}
\end{figure}

Here, $\overbar{Y_{i}}$ is the top-k candidates for replacing statement $X_i$. It is generated by our statement prediction model with the input skipgram $X_{i-1}, X_{i+1}$. Similarly, $\overbar{Y_{i, i+1}}$ is the top-k candidate statements for insertion between the statements $X_i$ and $X_{i+1}$, these candidates are generated by the statement prediction model with the input skipgram $X_i$ and $X_{i+1}$. The space of candidate programs $\bar{\mathcal{Y}}$ is shown in figure \ref{fig:sspace}, where each pair of arrows indicates a usage of the statement prediction model, and each row of colored bars represent a list of candidate statements.

To select a candidate program out of this search space, we choose 1 candidate statement out of every list of candidate statements, and concatenate the chosen statements together to form a candidate program. Since there are $k$ choices for each list, there are a total of $k^{2n+1}$ programs in our search space. Clearly a naive enumeration over the search space is infeasible, but each candidate statements also come with probabilities of generating that candidate, which we can use to construct a probability distribution $Pr(\mathcal{Y} | \mathcal{X})$, the probability of generating the candidate program given the original incorrect program. We can use this probability to guide our enumeration, trying more likely candidate programs first before attempting a less likely one. We define $Pr(\mathcal{Y} | \mathcal{X})$ as follows:

\begin{align}
\begin{split}
	& Pr(\mathcal{Y} | \mathcal{X}) \\
    & = Pr(Y_{0,1}, Y_{1} \dots Y_{n,n+1}| X_0 \dots X_{n+1}) \\
    & = \prod_{i}Pr(Y_i | X_{i-1}, X_{i+1}) \prod_{j}Pr(Y_{j,j+1} | X_{j}, X_{j+1}) \\
\end{split}
\end{align}

The probability of generating a candidate \emph{program} $\mathcal{Y}$ is factored into a product, each element of the product is the probability of generating a particular candidate \emph{statement} ,either for replacement or insertion, given by the statement prediction model. Notice that we made an independence assumption where each candidate statement is generated from the skipgram context in the incorrect program $\mathcal{X}$, rather than being dependent on the other candidate statements. An alternative way of generating a candidate program would be to apply the statement transformations sequentially, so that subsequent transformations can depend on previous transformations. In practice though, that scheme is inefficient to run and does not yield much accuracy improvements.

\subsection*{Dealing with Unseen Partial Fragments}
Our model is trained on a corpus of fragments collected from correct programs. This corpus, however large, cannot be complete. As a result, there will be instances where our algorithm confronts a program fragment that is unfamiliar, because similar fragments do not exist in the training set. More specifically, there will be instances where a partial fragment is given to the statement prediction model, and all the top-k candidates generated are nonsensical. Here's are some nonsensical candidate statements generated by our model on an unfamiliar program fragment:

\begin{tikzpicture}[node distance = 0.0in]
\node(a)[anchor=west]{\ttfamily{}};
\node(b)[below=of a.south west, anchor=north west, draw, minimum width=1in]{
\begin{lstlisting}
if len ( x0 ) :
if len 1 [ ) :
if len 1 : - 1 ] :
if len 1 : - 1 : - 1 :
\end{lstlisting}
};
\end{tikzpicture}

As we can see, if we insist on replacing the original program statement by one of the nonsensical candidate statements, our algorithm would fail on a program with unfamiliar fragments. To remedy this, we artificially add in the original program statement as one of the candidate statements for replacement, with an artificial ``probability'' of $1.0$. Similarly, an artificial candidate for insertion by the empty statement is also introduced. The result of these artificial modifications is that or distribution over candidate programs $Pr(\mathcal{Y} | \mathcal{X})$ presented earlier becomes a likelihood rather than a real probability, but otherwise remains unchanged. 

A consequence of introducing these artificial modification is its effect on our enumeration: The program with the highest likelihood is the original program, and this likelihood gradually decreases as our enumeration explores different combination of changes, gradually modifying the original program to be more different.

\subsection*{The Enumeration Algorithm}

We now present the algorithm for enumerating the space of candidate programs. 

We'll denote all the candidate statements (both for Insertions and Replacements) as $y_{ij}$, where the subscript $i$ indicates the list of candidate statement this candidate is chosen from, and the subscript $j$ denote it is the jth candidate from the list. A bigger index of $j$ with the same index $i$ would yield a candidate statement with a smaller probability.

We denote the cost of the program $cost(prog)$ as the negative log-likelihood of the probability $Pr(\mathcal{Y} | \mathcal{X})$, where a bigger value correspond to a less likely program.

Let's define a $next$ function, that takes in a candidate $y_{jk}$, and return the next, more costly candidate from the candidates generated from beamsearch. $next(y_{j,k}) = y_{j,k+1}$.

We can now formally write our search algorithm over the space of possible corrections as follows:

\begin{lstlisting}
@Input: candidates yjk for each Yj
@Output: A correct program or Failure
programSearch({yjk}):
  prog = [y10, ..., yr0]
  budget = 0
  Q = Queue()
  Q.push(prog, cost(prog))
  while budget<B:
    prog = Q.pop()
    if correct(prog):
      return prog
    for j in 1..r:
      nxt_prog = [candidate 
         for candidate in prog]
      nxt_prog[j] = next(nxt_prog[j])
      Q.push(nxt_prog, cost(nxt_prog)
  return FAIL
\end{lstlisting}

This algorithm searches through the space of possible corrections, starting with the original program which has no changes, and gradually move away to a more expensive programs. It does so by popping the least costly program from a queue, and considering all possible ways of making this program more costly by trying out the next candidate statements at each of its possible sites. Since we use a queue to keep track of the least expensive program, the first program to be returned is guaranteed to be the most likely candidate program given the original incorrect program. The enumerate is bounded above by a budget $B$, in practice we use the value 5000, i.e. 5000 candidate programs are enumerated before the algorithm returns FAIL.

\section{Experiments}
\label{experiments}

We design the experiments to evaluate the overall accuracy of $sk\_p$, and perform a comparison study against autograder \cite{rishabh}, the state of the art approach in automatic feedback generation for MOOCs. We also provide a breakdown on the kind of programs corrected by $sk\_p$, validating our claim that syntactic errors are worth correcting, and that the fragment completion model works well even when confronted with a novel program. Finally, we attempt to give a crude upper-bound on the performance of our approach by implementing an exhaustive model, capable of memorizing all program fragments perfectly.

\subsection*{Data Set Generation}
To perform our experiments, the benchmarks need to be split into training, validation, and testing sets. For our method, the training is done exclusively on correct programs and testing is done exclusively on incorrect programs.

A naive scheme of splitting the data would be take all correct programs as training data, and take all incorrect programs as testing data. This naive scheme is misleading for the following reason: For each benchmark, the student submissions are ordered by submission time, an incorrect submission may be corrected by the same student sometime later in the data  set. Under this scheme, a model trained on a student's correct solution will be used to correct his own mistakes. 

To avoid this, we split the data into to parts: an ``early'' part consists of the first 90\% of the submission, and a ``late'' part consists of the remaining 10\% of the submission. The early part is filtered for correctness, and the correct programs are split 90\% - 10\% for training and validation. All incorrect programs in the early part are discarded. Similarly, all correct programs in the late part are discarded, and the incorrect programs become the testing set. Under this scheme, a model learned from past correct programs is used to correct future incorrect programs. 

\begin{table}[]
\centering
\begin{tabular}{|l|r|r|r|}
\hline
benchmarks      & \multicolumn{1}{l|}{training} & \multicolumn{1}{l|}{validation} & \multicolumn{1}{l|}{testing} \\ \hline
computeDeriv        & 1252                          & 140                             & 263                          \\ \hline
computeRoot         & 1617                          & 180                             & 84                           \\ \hline
evaluatePoly        & 2313                          & 258                             & 97                           \\ \hline
getAvailableLetters & 1051                          & 117                             & 59                           \\ \hline
getGuessedWord      & 315                           & 35                              & 127                          \\ \hline
isWordGuessed       & 902                           & 101                             & 109                          \\ \hline
oddTuples           & 8720                          & 969                             & 1981                         \\ \hline
\end{tabular}
\caption{Data breakdown for each benchmarks}
\label{tabl:split}
\end{table}

Table \ref{tabl:split} shows the data breakdown for our benchmarks. The most salient aspect of these numbers is that there is a considerable variance in the number of training data, ranging from 315 correct programs to 8720 correct programs. We will discuss its effect on the accuracy of our approach later.

\subsection*{Accuracy}

The accuracy of $sk\_p$ on the test set is shown in Table \ref{tabl:acc}. The average accuracy for all the benchmarks is 29\%, with individual accuracy as low as 13\% and as high as 49\%.

Of the 7 benchmarks assignments, autograder \cite{rishabh} provides correction models for 3 assignments which can generate good quality feedback in real-time (under 5 seconds per submission) at an average accuracy of 30\%. $sk\_p$ outperforms autograder with an average accuracy of 35\% on these 3 assignments, with an average correction time of 5.6 seconds. The result of this comparison is shown Table~\ref{tabl:compare}. 

Of these assignments, $sk\_p$ significantly out-performs autograder on 2 assignments while losing badly on the assignment computeDeriv. The discrepancy of accuracy highlights an important distinction: autograder use a well-tuned manual correction model, while $sk\_p$ learns appropriate fragment-completions from data. In the computeDeriv benchmark, a common mistake is the omission of a basecase, which can be fixed by inserting 2 statements together (an if statement followed by its body). This omission of basecase is explicitly encoded in the correction model for autograder, which handles it readily. On the other hand, since $sk\_p$ only inserts up to 1 statement between every pair of statements, it is inherently unable to correct this error. However, for the other 2 assignments, the correction model is not quite straight forward as adding a base case, and $sk\_p$ is able to achieve far better results by learning from data.

\begin{table}[]
\centering
\begin{tabular}{|r|r |r r|}
\hline
benchmark           & \#test & sk\_p & acc   \\ \hline
computeDeriv        & 263    & 33    & 0.125 \\ \hline
computeRoot         & 84     & 15    & 0.179 \\ \hline
evaluatePoly        & 97     & 47    & 0.485 \\ \hline
getAvailableLetters & 59     & 19    & 0.322 \\ \hline
getGuessedWord      & 127    & 33    & 0.260 \\ \hline
isWordGuessed       & 109    & 21    & 0.193 \\ \hline
oddTuples           & 1981   & 871   & 0.440 \\ \hline
\end{tabular}
\caption{Accuracy of sk\_p}
\label{tabl:acc}
\end{table}

\begin{table}[]
\centering
\begin{tabular}{|r|r|r r|r r|}
\hline
benchmark     & \#tests & sk\_p & auto & sk\_p acc & auto acc \\ \hline
computeDeriv & 263        & 33    & 131        & 0.125     & 0.498     \\ \hline
evaluatePoly & 97         & 47    & 19         & 0.485     & 0.196     \\ \hline
oddTuples    & 1981       & 871   & 383        & 0.440     & 0.193     \\ \hline
\end{tabular}
\caption{Comparison between sk\_p and autograder}
\label{tabl:compare}
\end{table}

\subsection*{Kinds of Corrections}
To understand what kinds of errors $sk\_p$ can fix, we provide a breakdown of different kinds of corrections on each of the benchmark assignments, shown in Table~\ref{tabl:breakdown}

\begin{table}[]
\centering
\begin{tabular}{|r|r|r|r|r|r|}
\hline
benchmark           & syn & sem & fresh & syn\%   & fresh\% \\ \hline
computeDeriv        & 4   & 29  & 29    & 12.12\% & 87.88\% \\ \hline
computeRoot         & 6   & 9   & 12    & 40.00\% & 80.00\% \\ \hline
getGuessedWord      & 17  & 30  & 16    & 36.17\% & 34.04\% \\ \hline
isWordGuessed       & 2   & 17  & 11    & 10.53\% & 57.89\% \\ \hline
getAvailableLetters & 0   & 33  & 13    & 0.00\%  & 39.39\% \\ \hline
evaluatePoly        & 1   & 20  & 12    & 4.76\%  & 57.14\% \\ \hline
oddTuples           & 195 & 676 & 117   & 22.39\% & 13.43\% \\ \hline
\end{tabular}
\caption{Breakdown of different kinds of corrections}
\label{tabl:breakdown}
\end{table}

For these benchmarks, syntax errors constitute 18\% of all the errors fixed by $sk\_p$, and on some benchmark accounts for as much as 40\% of the corrections. This highlights the importance of handling syntactic errors and an advantage of our approach versus a symbolic corrector. A correction is marked as \emph{fresh} when $sk\_p$ generates a correct candidate program that's not one of the programs used during training. On average, 53\% of the corrections are fresh. This confers the advantage of the fragment completion model: The errors are fixed locally, without considering the rest of the programs, fixing only the program fragment which the model knows about and leaving the rest alone. As a result, our approach can work with novel programs as long as it has a particular program fragment that is familiar, rather than requiring the entire program to be familiar.

\subsection*{Efficacy of Neural Network}
Our fragment completion model is implemented with a neural network, which learns a function mapping from the partial fragment to a distribution on the missing statements. How well can this function be learned is largely a function of 2 variables: How big is the training data (the size of the training set) and how many different patterns is there to be learned (the number of unique fragments in the training set). 

We test how well does our model learn these fragments by implementing an \emph{exhaustive} model which memorizes all program fragments during training explicitly. We found that the neural network model only performs better than the exhaustive model when there is a relatively large number of training data relative to the number of fragments need to be learned.

Formally, the exhaustive model represents the \emph{empirical distribution} of the missing statement, conditioned on the partial fragment context. This probability is given in Equation \ref{eqn:empdis}. Here, $count(X,Y,X')$ denotes the total number of occurrences of the program fragment $X,Y,X'$ in the training corpus, and $count(X, X')$ denotes the total number of occurrences of the partial fragment $X,X'$. Dividing these 2 counts yields the empirical distribution of the missing statement conditioned on the partial fragment.

\begin{align}
\label{eqn:empdis}
\begin{split}
	Pr(Y | X, X') = \frac{count(X,Y,X')}{count(X,X')}
\end{split}
\end{align}

One can use a dictionary to memorize the empirical distribution directly: The partial fragment $(X, X')$ becomes a key, and a list of potential missing statements, along with their probabilities becomes its value stored in the dictionary. 

To use the model in the task of fragment completion amounts to performing a dictionary look up: Given a partial fragment, look up candidate statement for this fragment already stored in the dictionary. Here is a catch: What if no partial fragment can be found in the dictionary that matches the given partial fragment? There are 2 approaches to this issue: By insisting on exact matches or by performing approximate matches. In the case of exact matches, a list of candidate statements is only returned when the partial fragment matches exactly with a key stored in the dictionary, and an empty list is returned otherwise. In the case of approximate matches, all the keys in the dictionary are compared with the input partial fragment, and the candidate statements from the ``closest'' key is returned.  In our experiment, we use the string-distance to measure the distance between keys. In the case of exact match, one risk the possibility of missing a correction when a similar partial fragment is stored in the dictionary; On the other hand, in the case of approximate match one risk giving too many bogus candidate statements even if no similar partial fragment are being stored. In the experiment both approaches are evaluated.

Table~\ref{tabl:exhaust} compares the performance of these different approaches. In the table, \#frag denotes the total number of fragments being stored in the exhaustive model, and the benchmarks are sorted by the ratio \#tr / \#fr, the number of training programs divided by the number of fragments within a particular benchmark. Conceptually, this ratio measures the easiness of training a neural network model: With more training data and less fragments to learn, the neural network should perform better. 

\begin{table}[]
\centering
\begin{tabular}{|r|r|r|r|r|r|}
\hline
benchmark           & \#frag & \#tr/\#fr & sk\_p & exact & aprox \\ \hline
computeDeriv        & 7444   & 0.168          & 0.125 & 0.240 & 0.183 \\ \hline
computeRoot         & 8366   & 0.193          & 0.179 & 0.250 & 0.238 \\ \hline
getGuessedW      & 1418   & 0.222          & 0.260 & 0.339 & 0.346 \\ \hline
isWordGuess       & 3156   & 0.286          & 0.193 & 0.514 & 0.339 \\ \hline
getAvailabl     & 2503   & 0.420          & 0.322 & 0.356 & 0.356 \\ \hline
evaluatePoly        & 3925   & 0.589          & 0.485 & 0.371 & 0.423 \\ \hline
oddTuples           & 5323   & 1.638          & 0.440 & 0.409 & 0.439 \\ \hline
\end{tabular}
\caption{Comparison of different models}
\label{tabl:exhaust}
\end{table}

Overall, $sk\_p$ has an average accuracy of 29\%, the exhaustive model which uses approximate matching performs better, at 33\%, and the exact model works best with an accuracy of 35\%. Therefore, for our particular set of benchmarks, explicitly memorizing all the fragments during training will yield better results. We see the worst performing benchmark, computeDeriv also has the hardest model to train, having many different fragments to learn from while only having a relatively few number of training programs. The accuracy of the neural network model increases as the model becomes easier to train, and on the two benchmarks where there are many correct programs to train from with relatively few fragments, the neural network model outperforms exhaustive memorization. The neural network is able to outperform the exhaustive model in two ways: First, it can learn a better distance metric, matching a given partial fragments to ones seen during training in a more structured way than pure string distance. Second, a neural network is capable of generating novel program statements not seen during training. The second case is rare, but we do show a novel correction in the showcase section below.

\subsection*{Correction Showcase}
Here we showcase some corrections our model is able to produce for the evaluatePoly benchmark, highlighting our model's capability at fixing different kinds of errors.

\paragraph{Removing a superfluous check} An extraneous if statement on line 4 is removed. 

\begin{lstlisting}
# incorrect
def evaluatePoly(poly, x):
    n = 0
    s = 0
    for e in poly:
        if e > 0:
            s += e*x**n
        n += 1
    return float(s)

# corrected
def evaluatePoly ( poly , x ) : 
  n = 0 
  s = 0 
  for e in poly : 
    s += e * x ** n 
    n += 1 
  return float ( s ) 
\end{lstlisting}

\paragraph{Fixing an operator} The incorrect program uses the wrong operator = for assignment on line 4 instead of the operator += for update.
\begin{lstlisting}
# incorrect
def evaluatePoly(poly, x):
    y = 0.0
    exp = 0
    for i in poly:
        y = i * x ** exp
        exp += 1
    return y
    
# corrected
def evaluatePoly ( poly , x ) : 
  y = 0.0 
  exp = 0 
  for i in poly : 
    y += i * x ** exp 
    exp += 1 
  return y 
\end{lstlisting}

\paragraph{Fixing an Extra Indentation} The incorrect program has its return statement mistakenly indented. Note this constitute as a \emph{semantic error} in our experiments, because the incorrect program parses correctly.
\begin{lstlisting}
# incorrect
def evaluatePoly(poly, x):
  ans =0.0
  for i in range (len(poly)):
     ans=ans+(poly[i]*(x**i))
     return ans

# corrected
def evaluatePoly ( poly , x ) : 
  ans = 0.0 
  for i in range ( len ( poly ) ) : 
    ans = ans + ( poly [ i ] * ( x ** i ) ) 
  return ans 
\end{lstlisting}

\paragraph{A local fix to a complicated program} Our algorithm is able to fix this rather complicated program by changing the return statement, which is unnecessarily complicated by the student, likely under the impression that the result needs to be rounded. Note the extraneous print statement is also removed.
\begin{lstlisting}
# incorrect
def evaluatePoly(poly, x):
  if len(poly) == 0:
    return 0.0
  sumx = 0
  step = len(poly)
  while step >= 1:
    step -= 1
    sumx = sumx + (poly[step]*(x**(step)))
    print sumx
  return round(float(sumx),1)

# corrected
def evaluatePoly ( poly , x ) : 
  if len ( poly ) == 0 : 
    return 0.0 
  sumx = 0 
  step = len ( poly ) 
  while step >= 1 : 
    step -= 1 
    sumx = sumx + ( poly [ step ] * ( x ** ( step ) ) ) 
  return sumx 

\end{lstlisting}

\paragraph{Suggestion of a novel program} This novel suggestion fixes the incorrect program by replacing the ``for'' loop with a very strange ``while'' loop, which only work because of the extraneous update function ``x2 += 1'' present in the incorrect program. This correction is obtained earlier during our work, where we've yet to map back the correct solution's variable back to their original names.
\begin{lstlisting}
# incorrect prog
def evaluatePoly ( x0 , x1 ) : 
  x2 = 0 
  x3 = 0.0 
  for x2 in range ( 0 , len ( x0 ) - 1 ) : 
    x3 = x0 [ x2 ] * x1 ** x2 + x3 
    x2 += 1 
  return x3 

# corrected prog 
def evaluatePoly ( x0 , x1 ) : 
  x2 = 0 
  x3 = 0.0 
  while x2 in range ( len ( x0 ) ) : 
    x3 = x0 [ x2 ] * x1 ** x2 + x3 
    x2 += 1 
  return x3 
\end{lstlisting}
\section{Related Works}
Of the works in data driven synthesis and automatic feedback generation for MOOCs, we found the following work most relevant to compare.

\paragraph{}
In \cite{fan}, the problem of automatic patch generation is considered. A ranking probability is learned from a corpus of correct patches, which is then used to patch an incorrect program by ranking a search space of possible patches, with the top-ranked patch that is also correct returned to the user. This work is most similar to our work in that they both consider the problem of program repair. However, in their work, the ranking probability is a \emph{discriminative} model, and the search space of the patches need to be defined separately. Also, all candidate patches in this search space needs to be ranked, because without evaluating the probability function, one does not know if a certain patch is likely or not. In contrast, we learn a \emph{generative} model, where the candidate programs are statements are generated according to its probability, which alleviates the issue of having to separately define a search space and enumerating over the entire search space.

\paragraph{}
In \cite{raychev}, the problem of code completion is investigated. The user leaves holes in the program for the system to complete, and a language model is used to suggest possible method calls to put in these holes. The suggestions are constraint by the semantic context of the hole, and only suggestions that meet these constraints are given to the user. Our work shows that in the context of MOOCs, a much simpler model that directly operates on the tokenized statement can deliver good results without the need of filtering the candidate statements through semantic context, but is sufficient to use these statements verbatim. Also, our work focus on program correction, where accuracy is measured on whether the entire program pass the test suite, rather than independently as accurate suggestions.

\paragraph{}
In \cite{raychev2}, the problem of code annotation and variable renaming is investigated. A graphical model is used to infer code properties such as variable names and type annotations from obfuscated and uncommented javascript programs. In this work, the original program is semantically equivalent to the annotated and renamed output program, whereas we focus on the task of program correction, which involves non-trivial restructuring of the incorrect program to change its semantics.

\paragraph{}
In \cite{clustering}, the problem of automatic feedback generation with clustering is explored. For a given MOOCs assignment, its correct programs are clustered and a canonical program elected as a representative for each cluster, forming a set of reference solutions. Given an incorrect student solution, it is matched by distance against the reference solutions, and the closest one is returned as the fix. Our work shows that often an incorrect solution has a correction that is only few changes away, different from any reference solutions. this is backed by the existence of a significant number of ``fresh'' corrections: a fix that results in a correct program which does not exist in the training set. This implies the clustering approach is mapping incorrect student solutions to an unlikely correct solution, when a correction that more closely resembles it could exist. In a sense, our work is an implicit form of ``local clustering'' without the manual burden of defining a distance metric. Similarly, in \cite{grading}, a distance metric between a incorrect student submission and a set of correct student solution is considered, but instead of using the distance to provide a correction, the distance is used to give a grade, with the grade inversely proportional to the distance.






\bibliographystyle{abbrvnat}

\begin{thebibliography}{}
\softraggedright

\bibitem[Rishabh et~al.(2013)]{rishabh}
Singh, Rishabh, Sumit Gulwani, and Armando Solar-Lezama. "Automated feedback generation for introductory programming assignments." ACM SIGPLAN Notices 48.6 (2013): 15-26.

\bibitem[Mikolov et~al.(2013)]{word2vec}
Mikolov, Tomas, et al. "Distributed representations of words and phrases and their compositionality." Advances in neural information processing systems. 2013.

\bibitem[Raychev et~al.(2014)]{raychev}
Raychev, Veselin, Martin Vechev, and Eran Yahav. "Code completion with statistical language models." ACM SIGPLAN Notices. Vol. 49. No. 6. ACM, 2014.

\bibitem[Long et~al.(2016)]{fan}Long, Fan, and Martin Rinard. "Automatic patch generation by learning correct code." ACM SIGPLAN Notices. Vol. 51. No. 1. ACM, 2016.


\bibitem[Mikolov et~al.(2013)]{word2vec}
Mikolov, Tomas, et al. "Distributed representations of words and phrases and their compositionality." Advances in neural information processing systems. 2013.

\bibitem[Pennington et~al.(2014)]{glove}
Pennington, Jeffrey, Richard Socher, and Christopher D. Manning. "Glove: Global Vectors for Word Representation." EMNLP. Vol. 14. 2014.

\bibitem[Kiros et~al.(2015)]{skip-thought}
Kiros, Ryan, et al. "Skip-thought vectors." Advances in Neural Information Processing Systems. 2015.

\bibitem[Gers et~al.(2000)]{lstm}
Gers, Felix A., Jürgen Schmidhuber, and Fred Cummins. "Learning to forget: Continual prediction with LSTM." Neural computation 12.10 (2000): 2451-2471.

\bibitem[Raychev et~al.(2015)]{raychev2}
Raychev, Veselin, Martin Vechev, and Andreas Krause. "Predicting program properties from big code." ACM SIGPLAN Notices. Vol. 50. No. 1. ACM, 2015.

\bibitem[Guwani et~al.(2016)]{clustering}
Gulwani, Sumit, Ivan Radicek, and Florian Zuleger. "Automated Clustering and Program Repair for Introductory Programming Assignments." arXiv preprint arXiv:1603.03165 (2016).

\bibitem[Singh et~al.(2016)]{grading}
Singh, Gursimran, Shashank Srikant, and Varun Aggarwal. "Question Independent Grading using Machine Learning: The Case of Computer Program Grading."

\bibitem[Cho et~al.(2014)]{seq2seq}
Cho, Kyunghyun, et al. "Learning phrase representations using RNN encoder-decoder for statistical machine translation." arXiv preprint arXiv:1406.1078 (2014).

\bibitem[Abadi et~al.(2016)]{flow}
Abadi, Martın, et al. "Tensorflow: Large-scale machine learning on heterogeneous distributed systems." arXiv preprint arXiv:1603.04467 (2016).
APA	


\bibitem[Carpenter.(2005)]{giga}
Carpenter, Bob. "Scaling high-order character language models to gigabytes." Proceedings of the Workshop on Software. Association for Computational Linguistics, 2005.

\bibitem[Lafferty et~al.(2001)]{beam}
Lafferty, John, Andrew McCallum, and Fernando Pereira. "Conditional random fields: Probabilistic models for segmenting and labeling sequence data." Proceedings of the eighteenth international conference on machine learning, ICML. Vol. 1. 2001.

\end{thebibliography}


\end{document}